\documentclass[prl, twocolumn, superscriptaddress, nofootinbib, nobibnotes]{revtex4-1}

\usepackage[english]{babel}
\usepackage{hyperref}

\usepackage[pdftex]{graphicx}
\usepackage{color}
\usepackage[toc,page]{appendix}
\usepackage[normalem]{ulem}      % for edition only -- can be removed in the final version

\usepackage{amsmath}
\usepackage{amssymb}
\usepackage{mathtools}
\usepackage{braket}

\usepackage[sort&compress]{natbib}

\newcommand{\un}[1]{\ensuremath{\,\mathrm{#1}}}
\renewcommand{\v}[1]{\ensuremath{\boldsymbol{#1}}}

\newcommand{\fig}[1]{Figure~\ref{fig:#1}}
\newcommand{\Fig}[1]{Figure~\ref{fig:#1}}

\newcommand{\eq}[1]{Equation~(\ref{eq:#1})}
\newcommand{\lr}[1]{\ensuremath{\left( #1 \right)}}

\newcommand{\I}{\mathrm{i}}
\renewcommand{\ap}{\alpha}

\newcommand{\Sg}{\Sigma}
\newcommand{\sg}{\sigma}

\newcommand{\ta}{\theta}
\newcommand{\abs}[1]{\left| #1 \right|}

\begin{document}

\title{Gradient-index electron optics in graphene pn junctions}

\author{Emmanuel Paredes-Rocha}
\email{eparedes@icf.unam.mx}
\affiliation{Instituto de Ciencias F\'isicas, Universidad Nacional Aut\'onoma de M\'exico,
  Cuernavaca, Mexico}

\author{Yonatan Betancur-Ocampo}
\email{ybetancur@icf.unam.mx}
\affiliation{Instituto de Ciencias F\'isicas, Universidad Nacional Aut\'onoma de M\'exico,
  Cuernavaca, Mexico}

\author{Nikodem Szpak}
\email{nikodem.szpak@uni-due.de}
\affiliation{Fakult\"at f\"ur Physik, Universit\"at Duisburg-Essen, Duisburg, Germany}

\author{Thomas Stegmann}
\email{stegmann@icf.unam.mx}
\affiliation{Instituto de Ciencias F\'isicas, Universidad Nacional Aut\'onoma de M\'exico,
  Cuernavaca, Mexico}

\begin{abstract}
  We investigate the electron transport in smooth graphene pn junctions,
  generated by gradually varying electrostatic potentials. The numerically
  calculated coherent current flow patterns can be understood largely in terms
  of semi-classical trajectories, equivalent to the ones obtained for light
  beams in a medium with a gradually changing refractive index. In smooth
  junctions, energetically forbidden regions emerge, which increase reflections
  and can generate pronounced interference patterns, for example, whispering
  gallery modes. The investigated devices do not only demonstrate the
  feasibility of the gradient-index electron optics in graphene pn junctions,
  such as Luneburg and Maxwell lenses, but may have also technological
  applications, for example, as electron beam splitters, focusers and
  waveguides. The semi-classical trajectories offer an efficient tool to
  estimate the current flow paths in such nano-electronic devices.
\end{abstract}

\maketitle

\section{Introduction}

The ballistic, beam-like propagation of electrons in graphene enables the
observation of optical-like phenomena in this material. This electron optics has
recently come into focus of research with several theoretical and experimental
contributions \cite{Lee2015, Cayssol2009, Williams2011, Brun2019, Cserti2007,
  Cheianov2007, Garcia-Pomar2008, Allain2011, Young2009, Chen2016, Huard2007,
  Neto2009, Bai2018, Zhou2019, Hu2018, Bai2017}. For example, it has been shown
that an electron beam, which hits the interface of a graphene pn junction,
behaves like a light beam at the interface of two materials with different
refractive indices. Hence, the reflection and refraction of the electrons
follows a generalized version of Snell's law, where the refractive indices are
determined by the electrostatic potential in the p and n region of the junction
\cite{Katsnelson2006, Lee2015, Chen2016, Young2009, Cheianov2007, Cheianov2006,
  Stander2009}. Due to the special dispersion relation of graphene, negative
reflection can be observed, a property that has been seen before only in the
light propagation in metamaterials \cite{Schurig2006, Pendry2000,
  Veselago1968}. Moreover, Klein tunneling -- the absence of backscattering at
normal incidence -- is observed, which can be attributed to the pseudo-spin
conservation of the electrons in graphene. It has also been shown that an
electron beam in graphene can be deflected by means of elastic deformations that
induce a strong pseudo-magnetic field \cite{Levy2010, Guinea2010, Low2010,
  Stegmann2016, Amorim2016, Naumis2017}.

The possibility to manipulate electron beams in graphene by means of pn
junctions or elastic deformations has lead to various proposals for
nano-electronic devices, such as Veselago lenses \cite{Heinisch2013,
  Betancur-Ocampo2018, Betancur-Ocampo2018a, Mu2011, Cheianov2007, Brun2019,
  Peterfalvi2009, Prabhakar2019, Garg2014}, electron fiber optics
\cite{Williams2011, Jiang2017}, interferometers \cite{Mrenca-Kolasinska2016,
  Khan2014}, valley beam splitters \cite{Garcia-Pomar2008, Stegmann2018,
  Zhai2011, Zhai2018, Grujic2014, Schaibley2016, Settnes2017, Settnes2016,
  Milovanovic2016, Zhai2018, Carrillo-Bastos2018, Carrillo-Bastos2016,
  Rickhaus2015, Stegmann2016, Wang2018}, collimators \cite{Park2008, Liu2017},
switches \cite{Sajjad2011}, reflectors \cite{Gunlycke2014, Graef2019},
transistors \cite{Cayssol2009, Wang2019, Wilmart2014}, and Dirac fermions
microscopes \cite{Boggild2017}. Electron optics has been extended recently from
graphene to other materials, such as phosphorene where negative reflection has
been predicted \cite{Betancur-Ocampo2019, Betancur-Ocampo2020}, non-coplanar
refraction and Veselago lenses in Weyl semi-metals \cite{Hills2017, Yang2019,
  Yang2020, Lu2018}, anomalous caustics in borophene pn junctions
\cite{Zhang2019}, and super-diverging lenses in Dirac materials
\cite{Betancur-Ocampo2018}.

Most of the work on electron optics in graphene pn junctions involves interfaces
where the electrostatic potential (and hence the refractive index) changes
abruptly \cite{Cohnitz2016, Wu2014, Liao2013, Heinisch2013, Garg2014,
  Allain2011}. Recent experiments have demonstrated that such abrupt junctions
can indeed be realized \cite{Bai2018}. Pn junctions with a smoothly changing
electrostatic potential are often regarded as disadvantageous, because they
induce an energetically forbidden region and hence, reduce the
transmission. Nevertheless, one can also take advantage of the reduced
transmission to construct quantum dots based on smooth circular pn
junctions. These junctions show interesting physical properties like Mie
scattering \cite{Heinisch2013} and whispering gallery modes
\cite{Zhao2015}. They have been realized recently in experiments \cite{Brun2019,
  Brun2020}.

A smoothly changing electrostatic potential can be understood as a smoothly
changing refractive index that establishes gradient-index optics. In this paper,
we investigate to which extent graphene pn junctions show gradient-index optics
phenomena. We will study straight pn junctions as well as circular junctions
which have received little attention so far. On the one hand side, we will
calculate numerically the current flow in these systems, applying the
non-equilibrium Green's function method to the tight-binding model. On the other
hand side, using the geometric optics approximation, we will determine the
semi-classical trajectories of the electron beams. Comparing both approaches we
will show that they agree in a wide regime of parameters. Nevertheless,
discrepancies emerge which can be explained by the interference of electron
waves. These wave effects increase for smooth circular junctions as they can
partially confine the electron beam.
 
\vspace{1.5cm}

\section{System \& Methods}

\subsection{Graphene pn junctions}

We model the electronic structure of graphene by the tight-binding Hamiltonian
\begin{equation}
  \label{eq:1}
  H= -t \sum_{i,j} \ket{i^A}\bra{j^B} +\text{H.c.}
 \end{equation}
 which describes well the electron transport at low energies. The
 $\ket{i^{A/B}}$ indicate the atomic states localized on the carbon atoms at
 positions $\v r_i$ on the sub-lattices A and B, respectively. The sum runs over
 nearest neighboring atoms, which are separated by a distance of
 $a=0.142 \un{nm}$ and coupled with the energy $t= 2.8 \un{eV}.$ A plane-wave
 ansatz leads at low energies to the continuous Dirac Hamiltonian
\begin{equation}
  \label{eq:2}
  H_{\text{Di}}^\pm(\v k) = \hbar v_F \v \sg_\pm \cdot \v k,
\end{equation}
at the two Dirac points $\v K_\pm= \lr{0,\pm 4 \pi/(3\sqrt{3}a)}$ at the edges
of the Brillouin zone. The wavevector $\v k$ is measured with respect to these
Dirac points. We define $\hbar v_F= 3 a t/2$ and
$\v \sg_-= \v \sg_+^*= (\sg_1, \sg_2)$. The valley degree of freedom of the
electrons in graphene, which may be used for a new kind of electronics called
valleytronics \cite{Schaibley2016, Settnes2016, Settnes2017, Milovanovic2016,
  Carrillo-Bastos2016, Carrillo-Bastos2018, Zhai2018}, will not be relevant in
the present work, because the considered pn junctions do not affect it. The
Dirac Hamiltonian leads to the well-known conical energy bands of graphene
\begin{equation}
  \label{eq:3}
  E(\v k)=s\hbar v_F \abs{\v k},
\end{equation}
where $s= \text{sign}(E)= \pm 1$ is the band index.

\begin{figure}[t]
  \centering
  \includegraphics[scale=0.35]{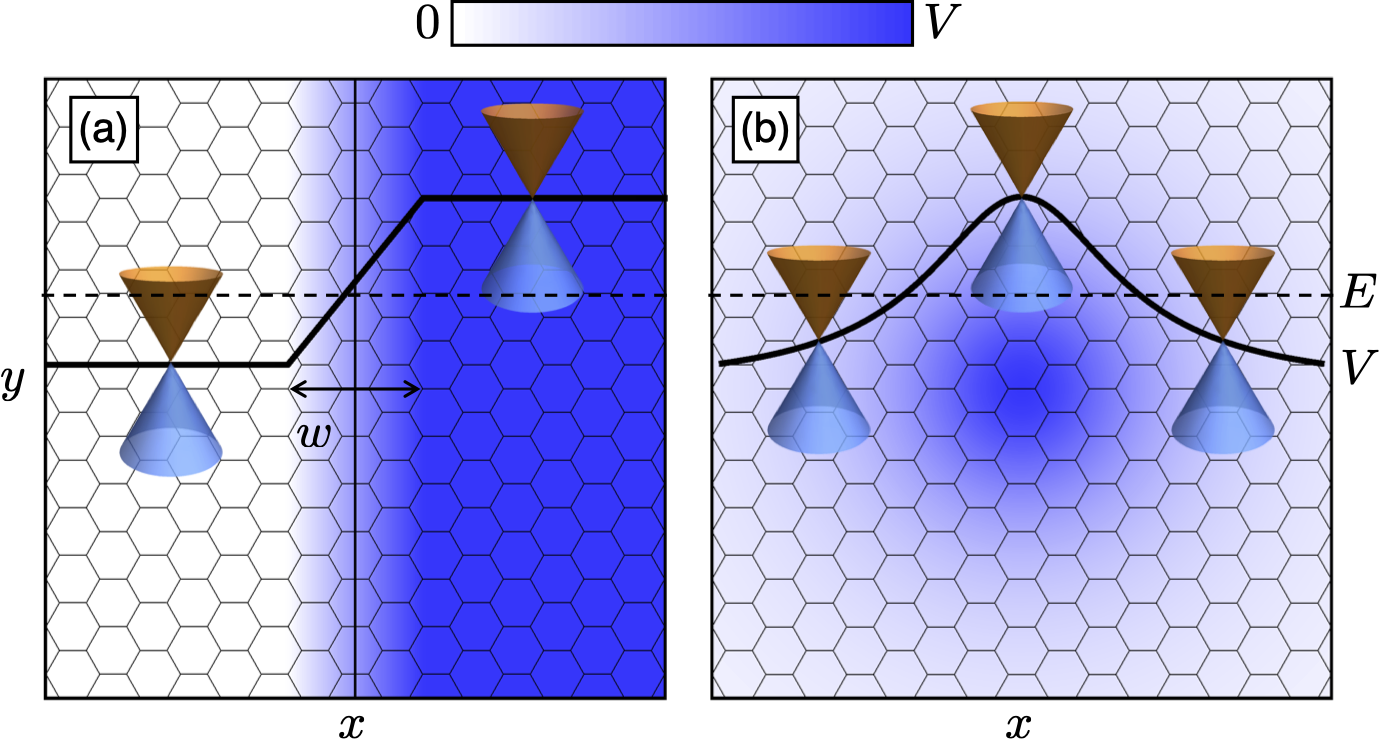}
  \caption{Straight (a) and circular (b) graphene pn junctions. The bluish color
    shading shows the electrostatic potential within the graphene ribbon. Both
    junctions have a smooth interface that separates the two regions of
    different doping. The profile of the electrostatic potential is indicated by
    the solid black lines, which shifts locally the conical energy bands of
    graphene. This causes that the electrons at energy $E$ (black dashed lines)
    go from the conduction band (orange cone) to the valence band (light blue
    cone).}
  \label{fig:1}
\end{figure}

A graphene pn junction is constituted by regions of different doping, see
\fig{1}. Such regions can be generated by metallic gates that induce an
electrostatic potential, $V(\v r)$, in the continuous space representation or
$V= \sum_i V_i \ket{i} \bra{i}$ in the tight-binding approach. This potential
shifts the energy bands of graphene and hence, generates a junction. A pn
junction is generated if the electrons go from one band to another through
interband tunneling, while in a pp' or nn' junction intraband tunneling takes
place within the valence or conduction band, respectively. In the following, we
will concentrate on straight pn junctions (\fig{1}~(a)), generated by the
electrostatic potential
\begin{equation}
  \label{eq:junc1}
  V_{\text{lin}}(\v r)=
  \begin{cases}
    0 &\text{if} \quad x \leq -w/2 \\
    \lr{\frac{x}{w}+\frac{1}{2}} V &\text{if} \quad \abs{x} < w/2\\
    V &\text{if} \quad x \geq w/2
 \end{cases}
\end{equation}
and circular junction (\fig{1}~(b)) with the potential
\begin{equation}
  \label{eq:junc2}
  V_{\text{cir}}( \v r)= \frac{V}{1+(r/r_0)^{\ap}}.
\end{equation}
The smoothness\footnote{Note that mathematical smoothness is not relevant here.}
of the pn junctions is controlled by the parameters $w$ and $\ap$. Note that
circular pn junctions can be also understood as pnp junctions because the
electrostatic potential is vanishing for $r \to \infty$.

\subsection{Semi-classical trajectories}

Within semi-classical theory, the propagation of the electron wave functions is
approximated by the propagation of point particles. In order to apply this
approximation to the quantum system described by the Dirac Hamiltonian \eq{2},
we use the eikonal approximation and the methods developed in \cite{Stegmann2016}
to obtain relativistic trajectories (geodesics) for massless particles coupled
to the electric potential $V(\v r)$
\begin{equation}
  \frac{d\v p }{dt} \equiv \frac{d}{dt} \left( \frac{(E-V(\v r))}{v_F^2} \frac{d\v r}{dt} \right)= - \nabla V(\v r),
\end{equation}
where the momentum vector $\v p(t)$ along the trajectory $\v r(t)$ satisfies the
dispersion relation $(E-V(\v r))^2 = v_F^2 \v p^2$. This approach is discussed
in detail in our previous works and applied successfully to understand the
current flow paths in elastically deformed graphene \cite{Stegmann2016,
  Stegmann2018} .

In order to get additional insight, the dynamics can be reformulated by means of
the classical pseudo-relativistic Hamiltonian
\begin{equation}
  \label{eq:4}
  H_{\text{sc}}= s(\v r)\,v_F\, \abs{\v p} + V(\v r),
\end{equation}
where $ s(\v r)=\text{sign}(E-V(\v r)) $ is the band-index and
$ \abs{\v p} = \sqrt{p_x^2 + p_y^2}= \sqrt{p^2_r + p_{\ta}^2/r^2} $ is the
momentum in Cartesian and polar coordinates, respectively. The trajectories of
the ballistic electrons described by this Hamiltonian can be related to optical
rays in an artificial medium with the refractive index
\begin{equation}
  \label{eq:5}
  n(\v r) \equiv \frac{E - V(\v r)}{v_F}.
\end{equation}
Taking into account that the electrostatic potential $V(\v r)$ can change
smoothly (on the length scale defined by the Fermi wavelength of the electrons),
we obtain in this way a gradient-index medium.

The equations of motion in a straight junction are given by
\begin{eqnarray}
  \label{eq:6}
  \frac{d x}{dt} & = & \partial_{p_x}H =
                       \frac{s(x)\,v_F\,p_x}{\sqrt{p^2_x + p^2_y}} = \frac{v_F p_x}{n(x)} \nonumber \\
  \frac{d y}{dt} & = & \partial_{p_y}H =
                       \frac{s(x)\,v_F\,p_y}{\sqrt{p^2_x + p^2_y}} = \frac{v_F p_y}{n(x)}.
\end{eqnarray}
Eliminating the time dependency by dividing both expressions, we obtain for the
semi-classical trajectories
\begin{equation}
  \label{eq:7}
  y(x) = y_0 + \,p_y\int^x_{x_0}\frac{s(x') \, dx'}{\sqrt{n^2(x') -p_y^2}},
\end{equation}
where $\v r_0 = (x_0,y_0)$ is the initial position of the electron. A similar
analysis for the electron trajectories in circular junctions leads to
\begin{equation}
  \label{eq:8}
  \ta = \ta_0 + l\int^r_{r_0}\frac{s(r')\,dr'}{r'\sqrt{r'^2n^2(r') - l^2}},
\end{equation}
where $l \equiv p_\ta$ is the angular momentum. Note that the momentum component
$p_y$ is conserved in straight junctions due to the translational symmetry along
the $y$-axis (see \fig{1}~(a)), while the angular momentum $\equiv p_\ta$ is
conserved in circular junctions due to the rotational symmetry (see
\fig{1}~(b)).

\begin{figure}[t]
  \centering
  \includegraphics[scale=0.35]{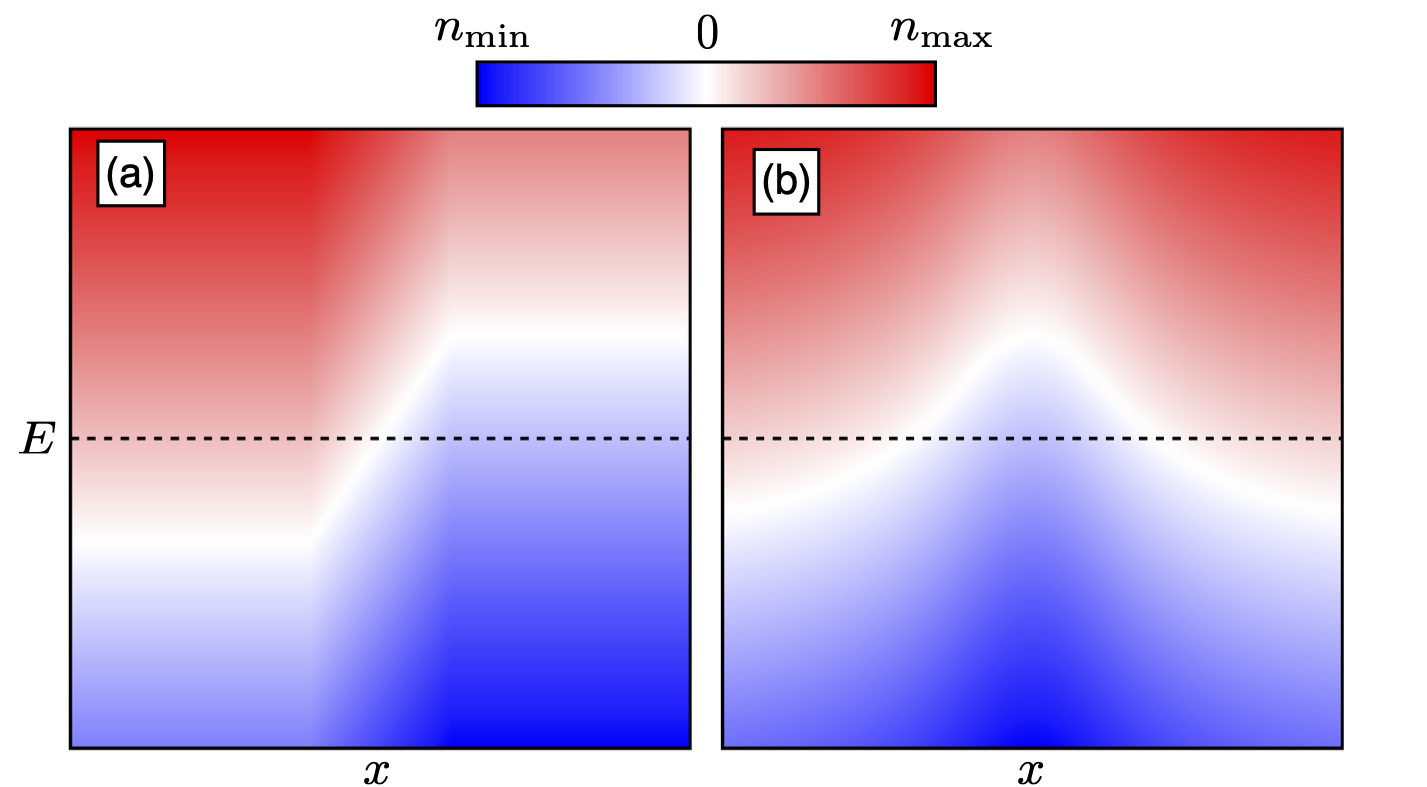}
  \caption{Refractive index $n$ in a straight (a) and circular (b) graphene pn
    junction as a function of the position $x$ (compare \fig{1}) and the
    electron energy $E$. Reddish colors indicate $n>0$, while bluish colors
    represent $n<0$. Note that for certain electron energies (dashed horizontal
    line) $n$ changes its sign in the narrow white region.}
  \label{fig:2}
\end{figure}

\begin{figure}[t]
  \centering
  \includegraphics[scale=0.35]{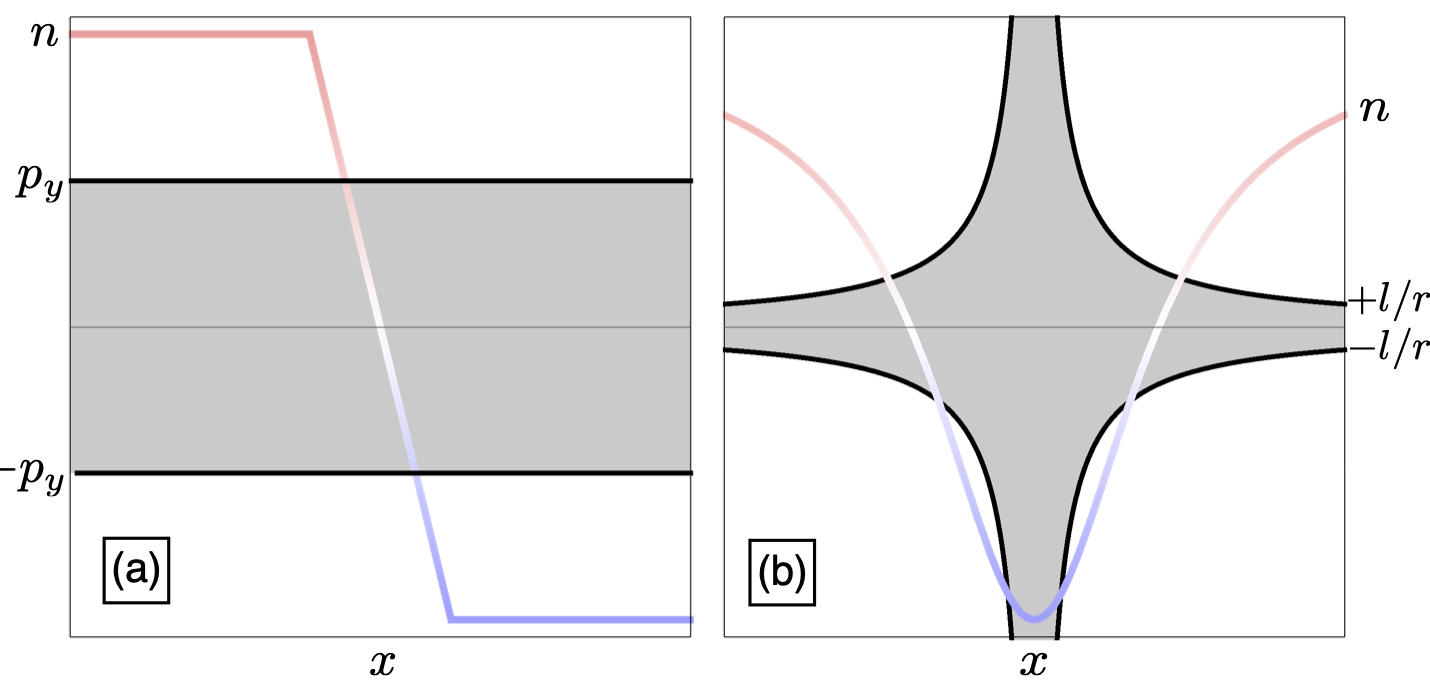}
  \caption{The reddish-bluish curve gives the refractive index $n$ at constant
    electron energy, see the dashed horizontal curves in \fig{2}. The black
    curves represent the momentum component $p_y$ and $l$ in straight (a) and
    circular (b) junctions, respectively.  When $n$ is in the gray shaded
    regions, the root in Eqs. \eqref{eq:7} and \eqref{eq:8} is imaginary and the
    electrons have to tunnel through a forbidden region.}
  \label{fig:3}
\end{figure}

These electron trajectories are identical to the ones obtained for light beams
in gradient-index optics \cite{Born1999}, apart from an important difference:
The refractive index $n(\v r)$ depends on the electron energy and changes its
sign when the electrons go from the conduction (n region) to the valence band (p
region), see \fig{2}. It is all negative for pp' and positive for nn'
junctions. Moreover, the square root in the denominator of Eqs. \eqref{eq:7} and
\eqref{eq:8} can become imaginary in certain regions of the system, which are
defined by the inequalities $|n(x)| \leq |p_y|$ and $|n(r)| \leq |l|/r$ for
straight and circular junctions, respectively. These forbidden regions are
indicated in \fig{3} by those ranges where the refractive index (bluish-reddish
curve) lies in the gray shaded regions.  While classically those regions cannot
be penetrated, quantum-mechanically the electrons can tunnel through the
forbidden regions.

As tunneling is largely suppressed in smooth junctions, the boundary of the
forbidden region defines the reflection zone for the beam. \Fig{3} explains that
the transmission decreases if the incidence of the electrons becomes more
oblique, because $p_y$ (or $l$) increases and, thereby, the size of the
forbidden region increases, too. In the same way, the transmission is perfect
for normal incidence, because $p_y=0$ (or $l=0$) and the forbidden region
disappears, which matches with the pseudo-spin conservation. The semi-classical
trajectories obtained from the geometrical optics are equivalent to the
relativistic geodesics in the classical region ($n>0$), but offer additional
techniques to deal with non-classical phenomena like tunneling though
classically forbidden regions. Note also that the refractive index in
\fig{2}~(b) is astonishingly similar to a measurement of the local resistance in
\cite{Brun2019}.

\subsection{The non-equilibrium Green's function method for the current flow}

The current flow in the graphene pn junction is calculated by means of the
non-equilibrium Green's function (NEGF) method. This quantum method is based on
the tight-binding Hamiltonian, \eq{1}. It does not rely on the approximations
made in the previous section to obtain the semi-classical trajectories and thus
allows us to verify their validity. As the NEGF method is discussed in detail in
various textbooks, see e.g. \cite{Datta1997, Datta2005}, we summarize here only
briefly the essential formulas.

The Green's function of the system is given by
\begin{equation}
  \label{eq:9}
  G(E)= \lr{E-H-V-\Sg}^{-1},
\end{equation}
where $E$ is the energy of the injected electrons, $H$ is the tight-binding
Hamiltonian, \eq{1}, and $V$ the electrostatic potential. In order to suppress
boundary effects and mimic an infinite system, we place a constant complex
potential $\Sg= -\I \sum_{i \in \text{edges}} \ket{i}\bra{i}$ at the edges of
the system, which should absorb the electrons.

The electrons are injected at the left edge of the system as plane waves
propagating towards the interface of the pn junction. Their momentum is
determined by the electron energy, \eq{3}, and the angle of injection
$\ta= \arctan{k_y/k_x}$. This injection is represented by the inscattering
function
\begin{equation}
  \label{eq:10}
  \Sg^{\text{in}}= \sum_{i,j \in \text{edge}} A(\v r_i) A(\v r_j)
  {\psi_{j}}^*(\v k) \psi_{i}(\v k) \ket{i} \bra{j},
\end{equation}
where the sum runs over all carbon atoms at the left system edge, see
\fig{1}. The $\psi_i(\v k)$ are the eigenstates of the Dirac Hamiltonian,
\eq{2},
\begin{equation}
  \label{eq:11}
  \psi_i(\v k){=}
  \begin{cases}
    c^{-} e^{\I(\v k{+}\v K^-)\v r_i} + c^{+} e^{\I(\v k{+}\v K^+)\v r_i} & i \in A,\\
    s\, c^{-} e^{\I(\v k{+}\v K^{-}) \v r_i + \I \phi} \,{-}\, s\, c^{+}
    e^{\I(\v k{+}\v K^{+}) \v r_i {-} i\phi} \hspace*{-2mm}& i \in B,
  \end{cases}
\end{equation}
where $\phi= \arg(\I k_x + k_y)$. The parameters $c^\pm$ control the occupation
of the two $\v K^\pm$ valleys. We consider the case in which both valleys are
fully mixed, i.e.  $c_\pm= \pm 1/2$. The function
\begin{equation}
  \label{eq:12}
  A(\v r)= e^{-(y - y_0)^2/d_0^2}
\end{equation}
gives the injected current beam a Gaussian profile. The parameters $y_0$ and
$d_0$ control the position and width of the beam.

Finally, the current flowing between the atoms at positions $\v r_i$ and
$ \v r_j$ is calculated by
\begin{equation}
  \label{eq:13}
  I_{ij} = \textrm{Im}(t\, G\, \Sigma^{\text{in}}\, G^{\dagger})_{ij}.
\end{equation}

Good agreement between the quantum current flow and the semi-classical
trajectories of geometric optics can be expected only in the specific parameter
regime, where the Fermi wavelength of the electrons $\lambda_F$ is much larger
than the interatomic distance $a$ but smaller than the system size
$\lr{L_x,L_y}$. Moreover, smooth changes of the electrostatic potential, as
sketched in \fig{1}, can be resolved only if the Fermi wavelength is shorter
than the spatial variations of the potential $\Delta_{x/y} V$. These conditions can
be summarized in the inequality
\begin{equation}
  \label{eq:15}
  a \: \ll \: \lambda_F= \frac{2\pi}{\abs{\v k}} \: < \Delta_{x/y} V \: \ll \: L_{x/y}.
\end{equation}
However, when the electrostatic potenial changes smoothly the effective electron
energy $E-V$, and hence the Fermi wavelength, will change which may lead to a
local violation of the inequality.

\section{Results}

\subsection{Straight graphene pn junctions}

% STEEP JUNCTION

\begin{figure}[t]
  \centering
  \includegraphics[scale=0.4]{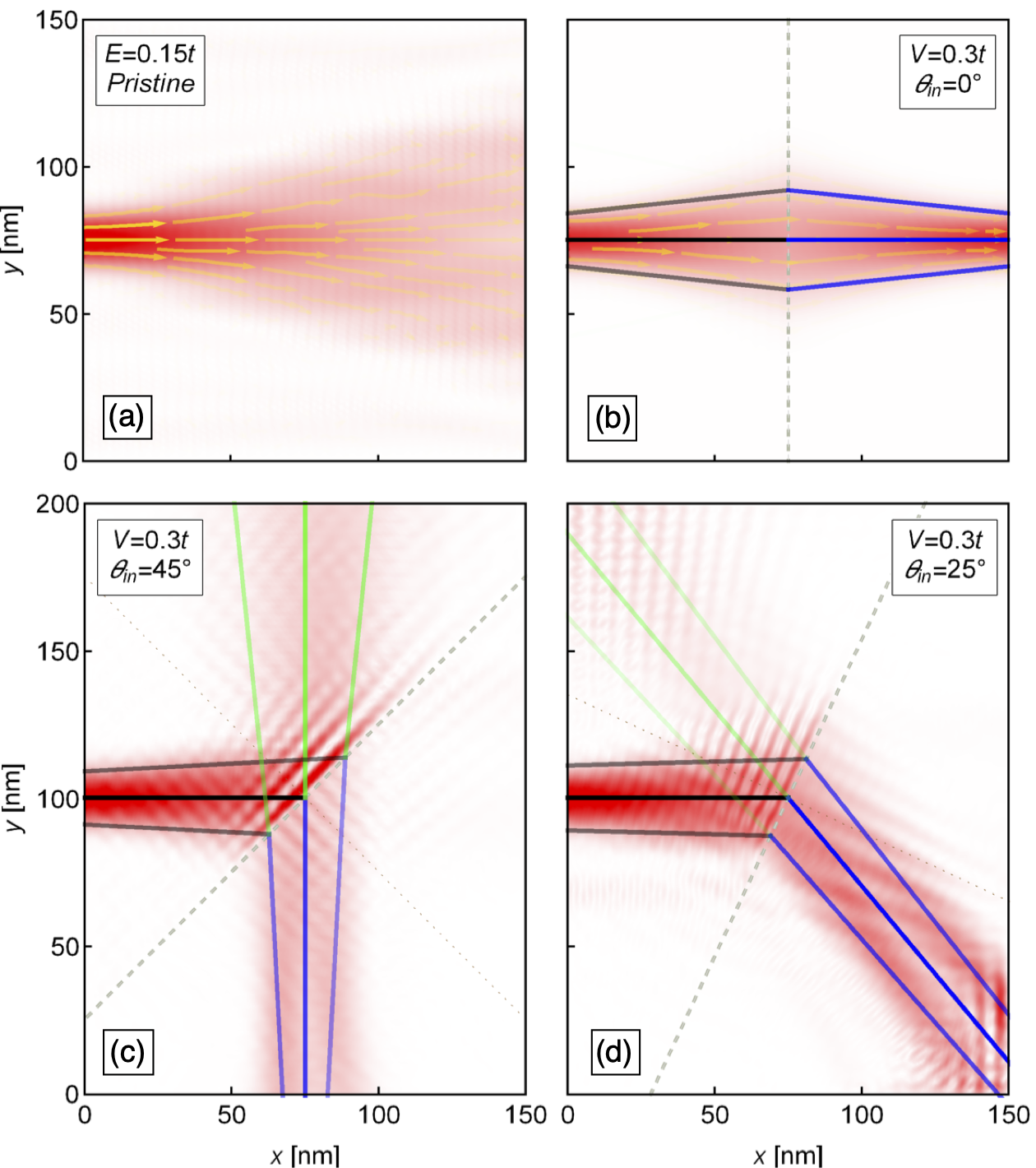}
  \caption{Current flow in a graphene nanoribbon in the absence (a) and in the
    presence (b,c,d) of an electrostatic potential which changes abruptly at the
    dashed line and hence, generates a pn junction. The current density is
    indicated by the red color shading, the current vector field by yellow
    arrows. The diverging electron beam in (a) is re-focused by the pn junction
    in (b). The semi-classical trajectories from \eq{7}, see the solid black,
    blue and green lines, agree with the current flow patterns calculated by
    means of the NEGF method. At the interface of the pn junction, the electron
    beam is split into a reflected and transmitted electron beam, in agreement
    with the generalized Snell's law, \eq{16}.}
  \label{fig:4}
\end{figure}

We begin our discussion by analyzing the current flow in a homogeneous graphene
nanoribbon with a size of about $150 \times 150 \un{nm}$ or larger\footnote{The
  accuracy of the used approximations increases with the system size.}. In
\fig{4}~(a), electrons are injected according to \eq{10} at the left ribbon edge
with energy $E=0.15t \approx 420 \un{meV}$ (corresponding to the Fermi
wavelength $ \lambda_F \approx 9 \un{nm} $) and a momentum vector parallel to
the horizontal $x$-axis. A beam-like propagation of the electrons with some
divergence due to diffraction is observed, which will enable us to compare the
numerically calculated current flow patterns with the semi-classical
trajectories.

\begin{figure}[t]
  \centering
  \includegraphics[scale=0.38]{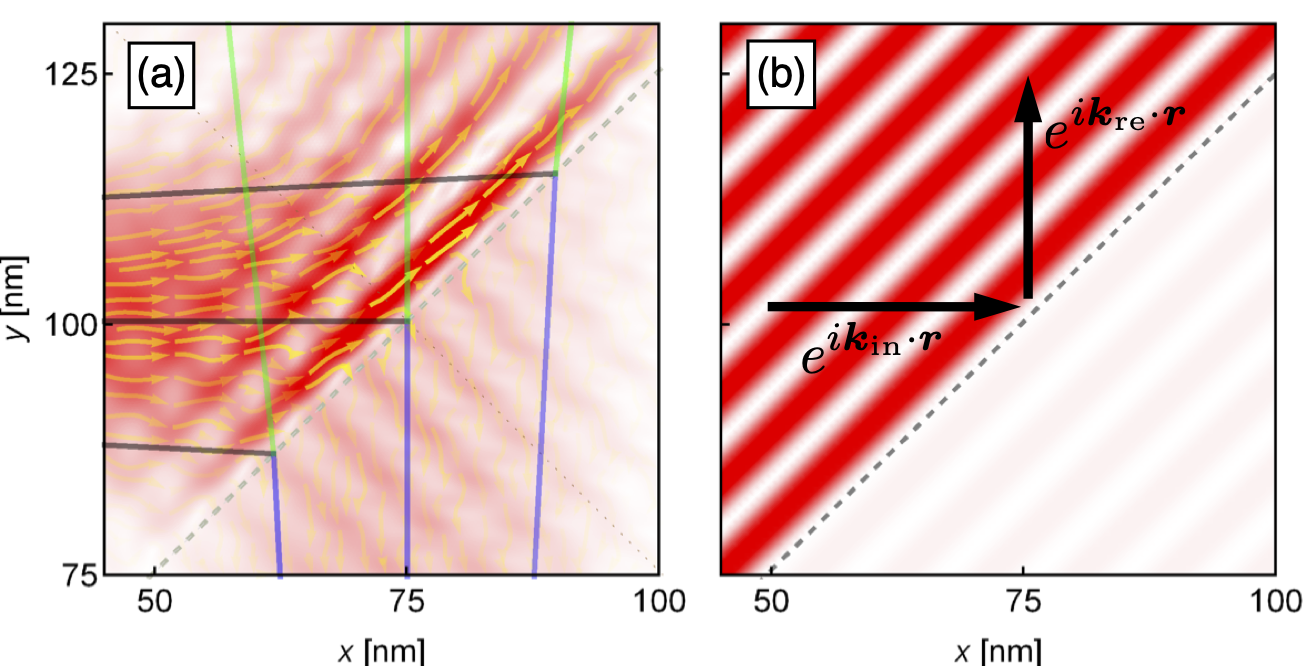}
  \caption{Current flow close to the interface of the graphene pn junction. (a)
    Zoom at the interface region of \fig{4}~(c). (b) Probability density
    generated by the superposition of the incoming and reflected electron beam,
    $\abs{e^{\I \v k_{\text{in}} \cdot \v r} + e^{\I \v k_{\text{re}} \cdot \v r}}^2$.
    The good agreement of (a) and (b) shows that the interference between the
    two electron beams generates the ripple pattern.}
  \label{fig:5}
\end{figure}

\begin{figure}[t]
  \centering
  \includegraphics[scale=0.4]{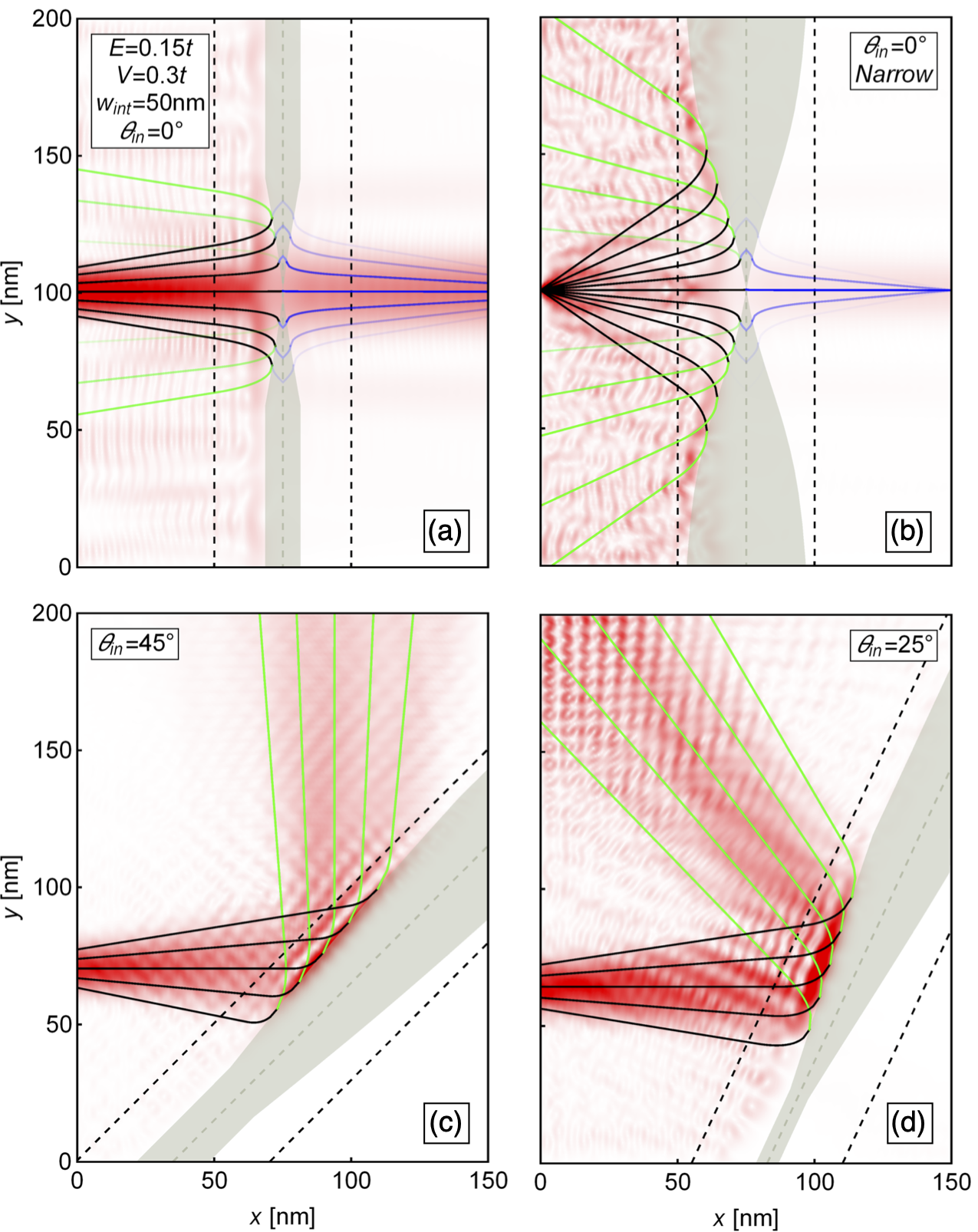}
  \caption{Current flow in a straight graphene pn junction with a smoothly
    changing profile ($w \sim 50 \un{nm}$). The width of the junction is
    indicated by dashed black lines (see \fig{1}~(a)). The points where the
    electrons go from the conduction to the valence band are indicated by a gray
    dashed line. The semi-classical trajectories (solid lines) agree well with
    the NEGF current density (red color shading). The forbidden region is
    indicated by the gray shaded region. At the edge of this region the current
    density accumulates and the semi-classical trajectories return. Note that
    the forbidden zone in the case of normal incidence (a,b) appears due to the
    diffraction of the electron beam, which has been determined on the basis of
    \fig{4}~(a).}
  \label{fig:6}
\end{figure}

In \fig{4}~(b) a straight pn junction is introduced by an electrostatic
potential which changes abruptly from zero to the constant value $V=2E$ at the
dashed line. We observe negative refraction at the interface of the pn junction
which re-focuses the electron beam. Moreover, Klein tunneling, i.e. the absence
of back-scattering at normal incidence, occurs. When the junction is
tilted\footnote{Note that due to the isotropy of graphene's electronic structure
  at low energies, tilting the junction is equivalent to injecting the electrons
  under a different angle.}  the electrons hit the interface no longer
orthogonally and the incident electron beam is split into a reflected and
refracted beam. The solid black, blue and green lines in \fig{4} are the
predicted semi-classical trajectories (\eq{7}) for the incident, transmitted and
reflected electron beams, respectively. They follow a generalized Snell's law,
\begin{equation}
  \label{eq:16}
  \ta_{\text{re}}= -\ta_{\text{in}} \quad \text{and} \quad
  \frac{\sin(\ta_{\text{in}})}{\sin(\ta_{\text{tr}})}=
  \frac{n_{\text{tr}}}{n_{\text{in}}}= \frac{E-V}{E},
\end{equation}
where $\ta_{\text{in/tr/re}}$ are the angles of incidence, transmission and
refraction, while $n_{\text{in/tr}}$ as defined in \eq{5} take the role of the
refractive indices in the n and p region, respectively. In general, these
trajectories agree very well with the numerical quantum calculations, see
\fig{4}~(b,c,d).  Electron optics in such straight pn junctions has been studied
largely before \cite{Allain2011, Bai2017, Boggild2017, Cayssol2009,
  Cheianov2006, Cheianov2007, Chen2016, Cohnitz2016, Garcia-Pomar2008, Hu2018,
  Huard2007, Lee2015, Low2009, Sajjad2011, Sajjad2012, Stander2009, Wang2019,
  Young2009, Zhou2019}. Here, we have confirmed the generalized Snell's law by
means of numerical quantum calculations of the current flow. Additionally, we
observe in \fig{4}~(c,d) a ripple pattern close to the interface of the pn
junction, magnified in \fig{5}~(a). In \fig{5}~(b) we show the superposition of
the incoming and reflected electron wave,
$ \abs{e^{\I \v k_{\text{in}} \cdot \v r} + e^{\I \v k_{\text{re}} \cdot \v
    r}}^2 $. The agreement of both figures confirms that the ripples are an
interference effect of the electron wave functions, which goes beyond
semi-classical trajectories. Similar ripple patterns can also be observed close
to the edges of the graphene nanoribbon. They can be explained in the same way
by the reflections at the system boundary. The absorption of the electrons at
the edges by the complex potential is not perfect and hence generates a small
reflected part which interferes with the incoming beam.

% SMOOTH JUNCTIONS

We continue our discussion with smooth graphene pn junctions, where the
electrostatic potential changes linearly over a width of
$w=350a \approx 50 \un{nm} \approx 5 \lambda_F$, see \fig{1}~(a). As shown in
\fig{6}, the semi-classical trajectories agree well with the current density
obtained by means of the NEGF method. The generalized Snell's law remains valid
for such smooth junctions. However, in comparison with the case of an abruptly
changing potential (see \fig{4}), a much larger part of the incident current is
reflected. This effect can be observed even for normal incidence, where Klein
tunneling takes place due to the diffraction of the electron beam which slightly
changes the propagation direction. The effect gets even more pronounced for
narrower electron beams for which the diffraction is stronger, compare
\fig{6}~(a) and (b). Such pn junctions can be used to generate narrow parallel
electron beams.

The increasing reflection can be understood by the rise of the forbidden zone
(see \fig{3}~(a)), which is indicated in \fig{6} by the gray shaded
regions. Moreover, an accumulation of the current density can be observed just
at the edge of the forbidden zone, which coincides with the point of return of
the semi-classical trajectories. Note that we have also sketched trajectories in
the forbidden regions by using the substitution $\v k \to \I \v k$, which
converts evanescent waves to propagating ones. Although the Fermi wavelength
diverges in this region and the geometric approximation might break down the
continuation of the trajectories to another classical region agrees again very
well with the quantum current.

% PP' and NN' junctions.

\begin{figure}[t]
  \centering
  \includegraphics[scale=0.4]{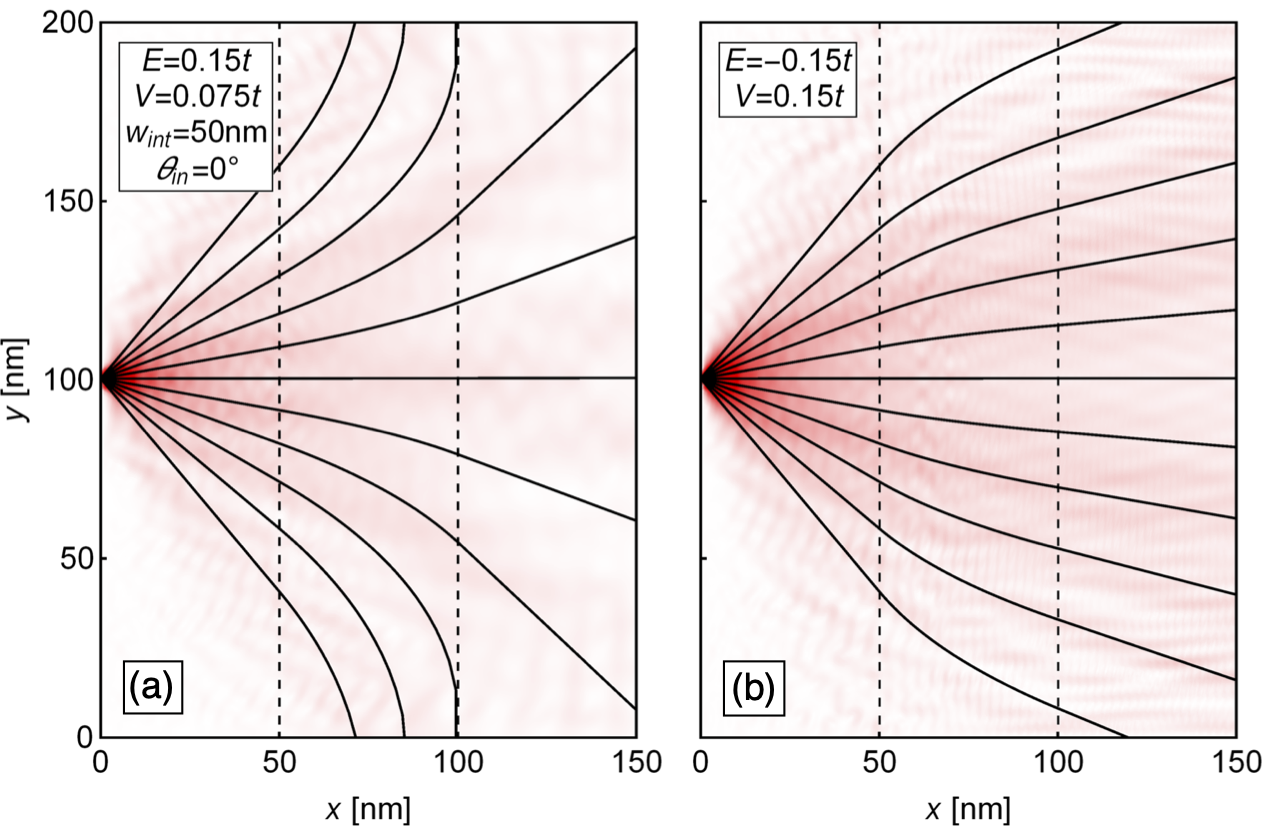}
  \caption{Electron optics in a smooth nn' (a) and pp' (b) junctions. The
    electrons are injected as a very narrow beam with strong diffraction. The
    current is transmitted largely through the junction in contrast to the pn
    junction, where the electrons are transmitted only at normal incidence
    (compare \fig{6}~(b)).}
  \label{fig:7}
\end{figure}

Until now, we have discussed only the case of pn junctions, where interband
tunneling occurs. In the case of nn' and pp' junctions, the electrons remain in
the same band and the current flow patterns change qualitatively. In \fig{7} a
narrow (and hence strongly diverging) electron beam is injected at the left
ribbon edge. In contrast to the pn junction (see \fig{6}~(b)) the current is
largely transmitted through the interface of the junction. Moreover, in the nn'
junction (\fig{7}~(a)) the divergence of the electron beam is enhanced, whereas
in the pp' junction (\fig{7}~(b)) it is reduced similarly to a converging
lens. Note that in nn' junctions a critical angle exists, beyond which total
reflection occurs, see the outermost trajectories in \fig{7} (a).

\vfill

\subsection{Circular graphene pn junctions}

\begin{figure}[t]
  \centering
  \includegraphics[scale=0.355]{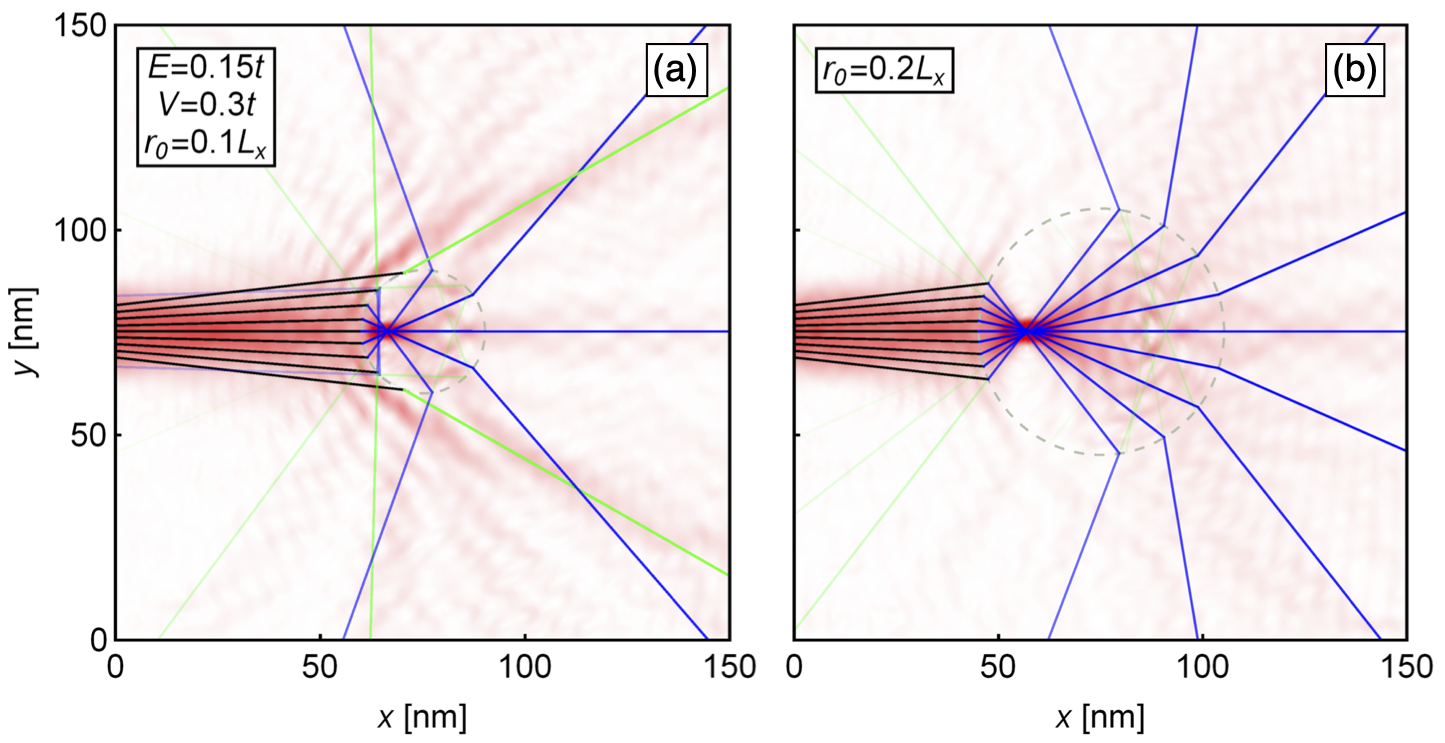}
  \caption{Current flow in a circular graphene pn junction. The electrostatic
    potential is changing abruptly at the dashed circle from zero to $V=2E$. A
    single focusing point emerges inside the junction, whose interface is
    denoted by the dashed circle. Outside the junction, the electrons are
    scattered divergently. The semi-classical trajectories agree well with the
    current flow inside the junction, but not that well outside due to
    interference of the incoming and reflected electron waves in a wide region.}
  \label{fig:8}
\end{figure}

\begin{figure}[t]
  \centering
  \includegraphics[scale=0.5]{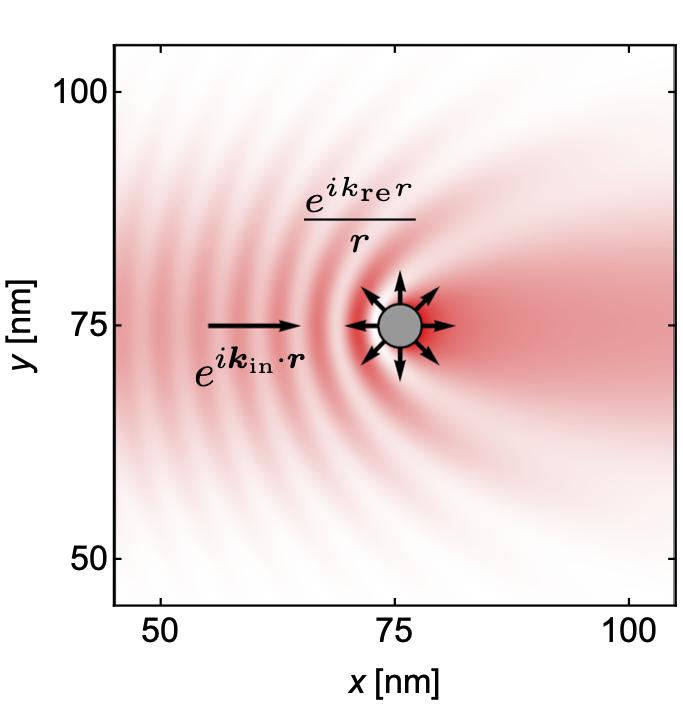}
  \caption{Interference of an incoming plane wave (with Gaussian profile, see
    \eq{12}) and a reflected circular electron beam,
    $\abs{A(\v r) e^{\I \v k_{\text{in}} \cdot \v r} + e^{\I k_{\text{re}} r}}^2$
    shows that the current pattern outside the junction is changed strongly by
    interference.}
  \label{fig:9}
\end{figure}

Let us consider now the circular graphene pn junctions. As in the case of
straight junctions, we begin with an abruptly changing electrostatic potential,
obtained by $\alpha \to \infty$ in \eq{junc2}. In \fig{8}, it can be observed
that a part of the current density is deflected around the pn junction while the
part that enters the junction is focused onto a single point
\cite{Cserti2007}. The semi-classical trajectories show a caustic inside the
junction in agreement with the focusing point observed in the current
density. However, outside the junction both approaches are much less in
line. This disagreement can be explained by the wave nature of the electrons
which leads in straight junctions to a ripple pattern at the interface of the
junction due to the interference between the incoming and reflected electron
waves. In circular junctions, the reflected electron wave is (approximately)
circular and hence interferes with the incident wave in a much larger
region. The resulting interference pattern shown in \fig{9} agrees qualitatively
with the observed current pattern and demonstrates the limitations of the
geometric optics in circular pn junctions. Additionally, we note in \fig{8} that
when the radius of the junction is reduced, the incidence of lateral electrons
becomes more grazing and a larger part of the electrons is deflected around the
junction.

% SMOOTH CIRCULAR JUNCTIONS

\begin{figure}[t]
  \centering
  \includegraphics[scale=0.55]{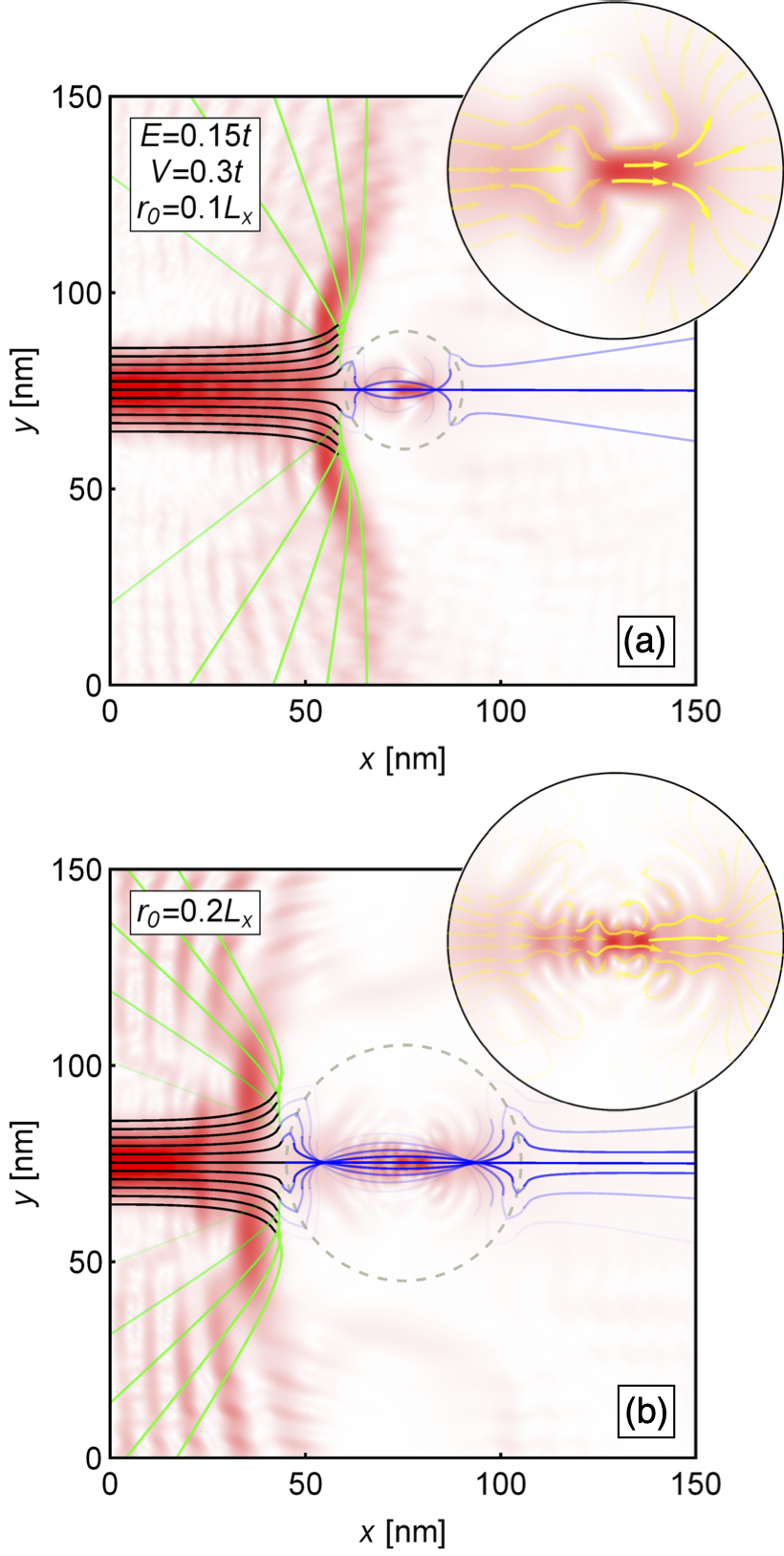}
  \caption{Gradient-index electron optics in smooth circular pn junctions
    ($\alpha=2$). The dashed circle indicates the isoline $V(r)=E$. The current
    flow pattern and the semi-classical trajectories agree roughly. The
    differences can be explained by the existence of an energetically forbidden
    region, which enhances reflections outside the junction as well as the
    confinement inside the junction and hence, leads to pronounced interference
    patterns. The interface of the forbidden region is indicated by the point
    where the trajectories change their color from black to green. The inset
    shows a zoom close to the center of the pn junction.}
  \label{fig:10}
\end{figure}

\begin{figure}[t]
  \centering
  \includegraphics[scale=0.55]{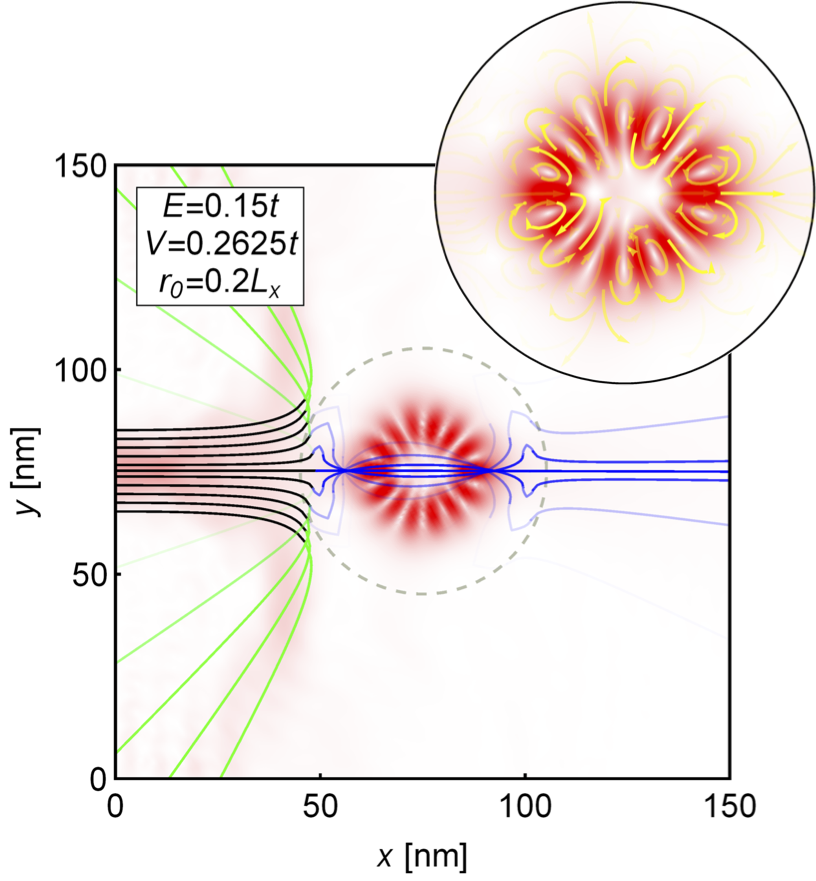}
  \caption{Gradient-index electron optics in smooth circular pn
    junctions. Strong interference patterns in the form of whispering gallery
    modes can be observed for specific parameters.}
  \label{fig:11}
\end{figure}

By smoothing the profile of the electrostatic potential ($\alpha=2$), we obtain
a circular junction with a gradually changing refractive index, which is
certainly the most appealing and novel device for gradient-index electron
optics. Such a device has been realized recently experimentally \cite{Brun2019,
  Brun2020}. Current flow patterns are shown in \fig{10} together with
semi-classical trajectories, which agree roughly with the quantum current. The
disagreement can be explained by strong interference due to an energetically
forbidden region through which the electrons have to tunnel. Therefore, a large
fraction of the incident electron beam is reflected, which causes a ripple
pattern that is much more pronounced than in abrupt circular junctions, see
\fig{8}. Similar to smooth straight junctions, see \fig{6}, an accumulation of
the current density can be observed at the edges the forbidden region, which is
indicated in the semi-classical trajectories by a color change from black to
green. The size of the forbidden region is minimal for the electrons of normal
incidence (see \fig{3}) and hence, those electrons enter and leave the junction
preferentially. Those electrons that enter the junction are strongly confined by
the Lorentz potential leading to various interference patterns. Moreover, for
some parameters, see \fig{11}, we can observe extremely pronounced interference
patterns inside the junction, such as whispering gallery modes \cite{Zhao2015,
  Jiang2017,Ghahari2017}. It is interesting to see that such modes can be
induced by an external electron beam.

\begin{figure}[t]
  \centering
  \includegraphics[scale=0.4]{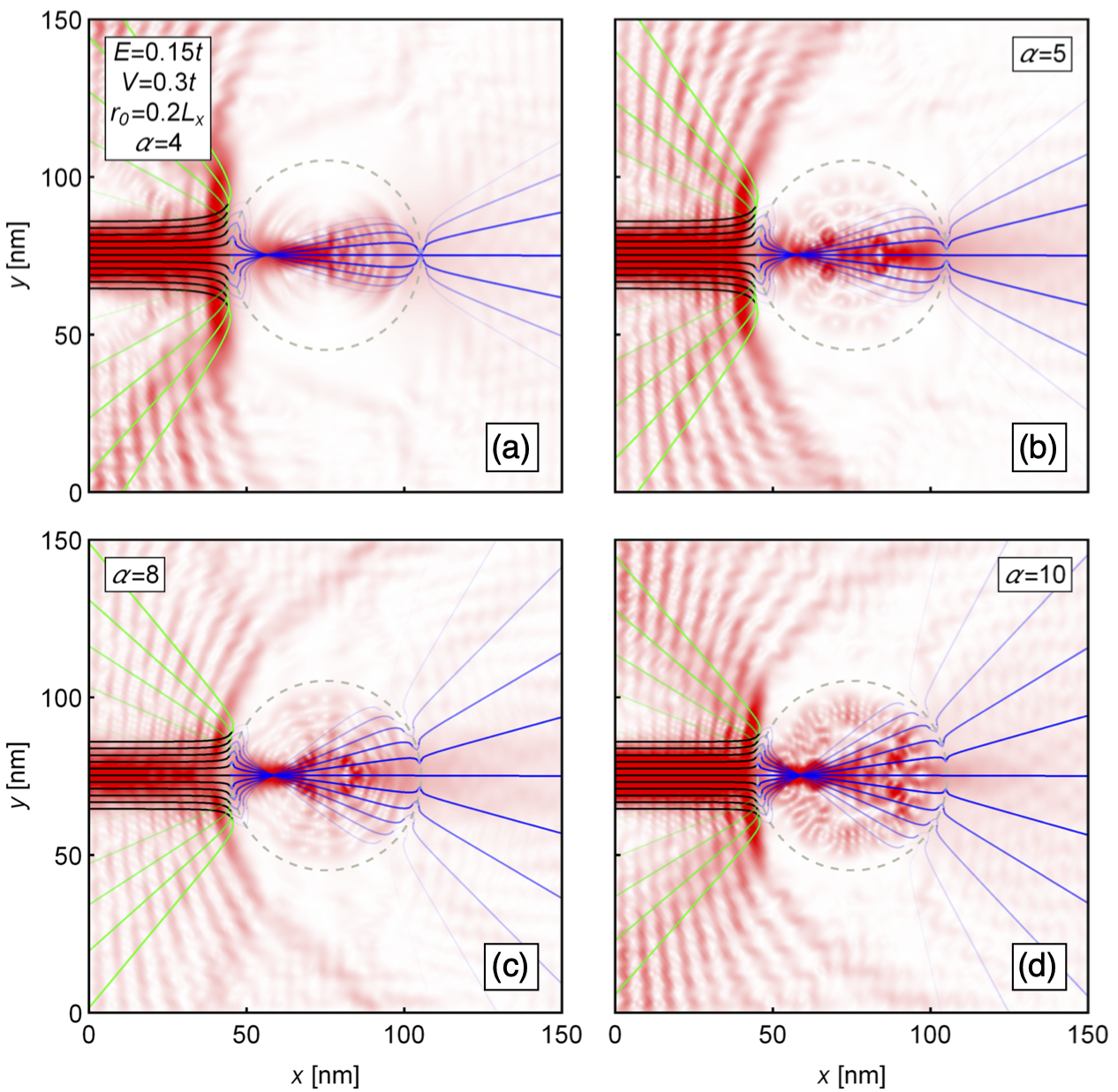}
  \caption{Current flow in circular pn junctions. When the parameter $\alpha$ is
    increased the junction changes more abruptly. The current density is
    dispersed more strongly and the focusing point moves from the center of the
    junction (see \fig{10}~(b)) towards the left (see \fig{8}~(b)). Moreover,
    distinct interference patterns can be observed inside the junction.}
  \label{fig:12}
\end{figure}

In \fig{12}, we study the transition from smooth to abrupt circular pn junctions
by increasing the parameter $\alpha$ in \eq{junc2}. When the junction profile
becomes more abrupt, the current density is dispersed more strongly and the
focusing point of the current moves away from the center of the junction
(see \fig{10}~(b)) towards the left edge of the junction (see
\fig{8}~(b)). Moreover, interference patterns are observed inside the junction,
which depend sensitively on the smoothness of the junction.

% NN' AND PP' JUNCTIONS

\begin{figure}[t]
  \centering \includegraphics[scale=0.4]{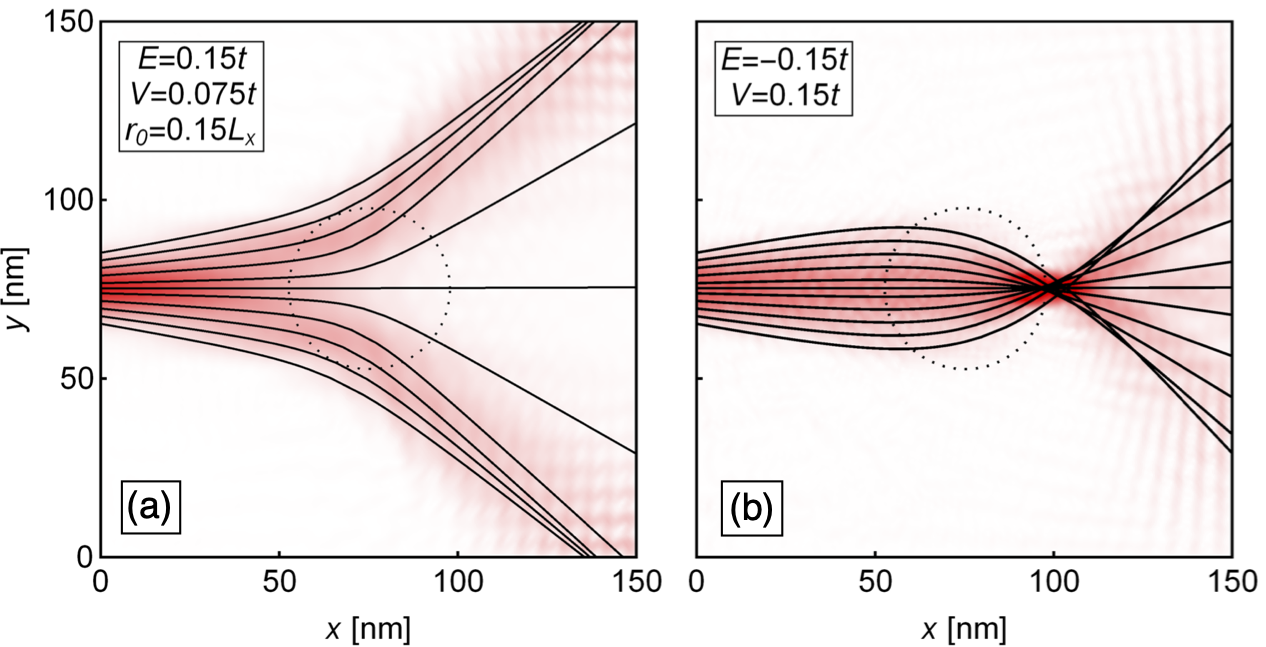}
  \caption{Current flow in smooth circular nn' (a) and pp' (b) junctions. The
    former represent efficient electron beam splitters, while the latter act as
    a converging lens.}
  \label{fig:13}
\end{figure}

\begin{figure}[t]
  \centering \includegraphics[scale=0.35]{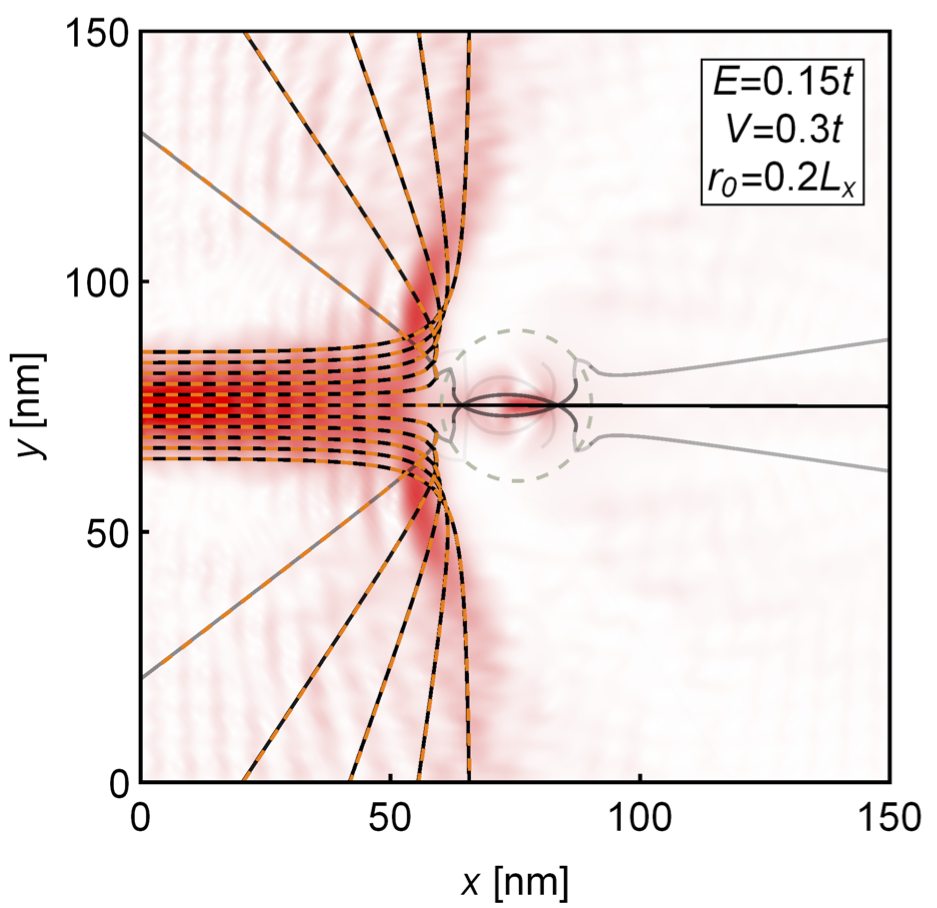}
  \caption{Current flow in a smooth circular pn junction. Semi-classical
    trajectories (see \eq{8}) are shown by solid-black curves, while the
    geodesics are given by the dashed-orange curves. Both approaches are
    equivalent for the reflected electrons. However, geodesics cannot be used to
    estimate the paths of the electrons that are transmitted through the
    junction.}
  \label{fig:14}
\end{figure}

When studying smooth circular nn' and pp' junctions, see \fig{13}, we observe
that the current flow patterns change qualitatively and more drastically in
comparison with straight junctions, see \fig{7}. In the nn' regime, the Lorentz
potential acts as a beam splitter, which separates even electrons with small
angular momentum. The semi-classical trajectories indicate that the Klein
tunneling persists for zero angular momentum (normal incidence). Once more, the
trajectories from \eq{8} and numerical calculations of the current density show
a remarkable agreement. Furthermore, a circular pp' junction behaves like a
converging lens, see \fig{13}~(b). The electron current flow is focused on a
single point behind the junction. In contrast with the pn regime, interference
patterns decrease, and therefore, pp' and nn' junctions are an ideal scenario to
realize gradient-index electron optics. It is important to note that the
refractive index defined by \eq{5} is fully positive in the nn' regime, while it
is all negative in the pp' regime. Moreover, we have demonstrated that the
gradient-index electron optics is in line with the principles of light optics
even for negative refraction, because ``the rays bend towards the region of
higher refractive index'' \cite{Born1999}. \Fig{14} shows the advantages
of the optical methods used here over the relativistic geodesics. Apart of the
classical regions where both approaches perfectly agree, we also see the
tunneling through a forbidden region, corresponding to tracing the evanescent
waves, and further propagation in a second classical region. In this way, we
obtain a much more complete picture and better agreement with the wave
dynamics.

\begin{figure}[t]
  \centering \includegraphics[scale=0.4]{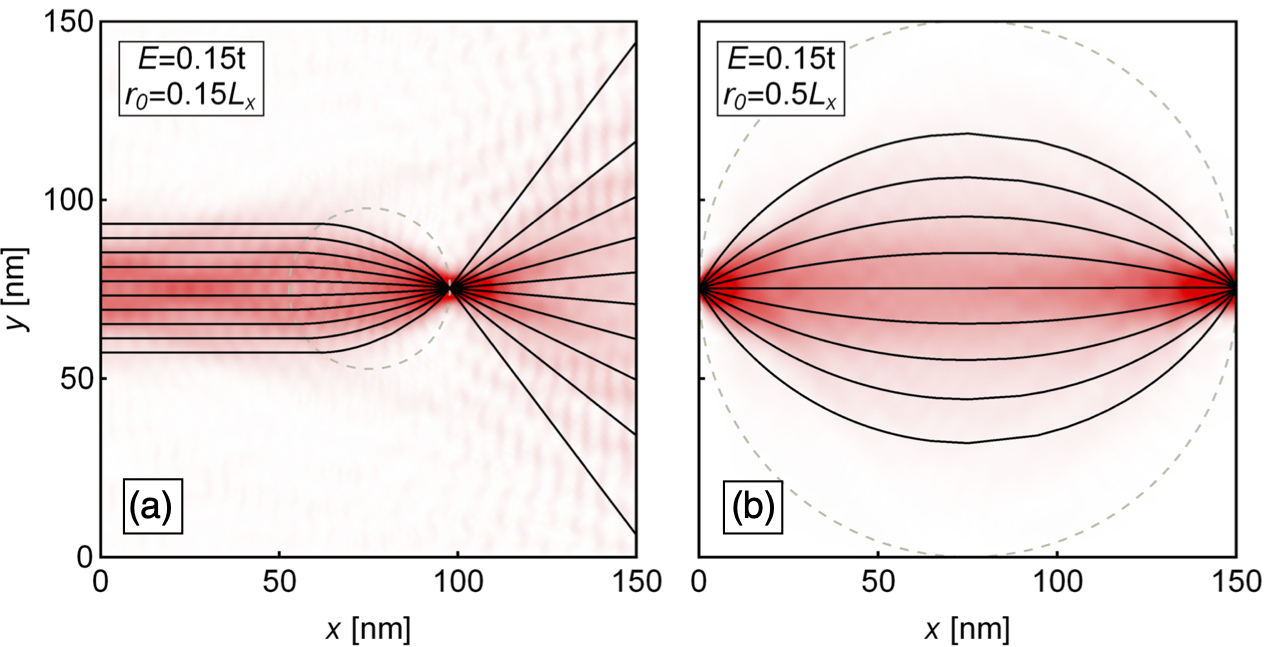}
  \caption{Current flow in graphene with an electrostatic potential that
    generates a Luneburg lens (a) and a Maxwell's fish-eye lens (b).}
  \label{fig:15}
\end{figure}

As a proof of principle, we also apply the developed techniques to the well
known Luneburg and Maxwell gradient-index lenses. The standard Luneburg lens
\cite{Luneberg1944} is known for its perfect focusing of parallel beams coming
from any direction and is described by
\begin{equation}
  n(\v r) = n_0
  \begin{cases}
    \sqrt{2 - (r/r_0)^2}\quad&\text{for}\quad r<r_0\\
    1 \quad&\text{otherwise}.\\
  \end{cases}
\end{equation}
In our situation, related to graphene, the interesting parameter is the
potential $V(\v r)$ which follows from \eq{5} and $n_0 = E/v_F$.  The Maxwell's
fish-eye lens \cite{Maxwell1854} has all pairs of focusing points on a circle
and is generated by the refractive index
\begin{equation}
 n(\v r) =\frac{n_0}{1+ \lr{r/r_0}^2},  
 \end{equation}
 which is closely related to the previously used potential in \eq{junc2}. The
 functionality of these electron optical devices in graphene in demonstrated in
 \fig{15}.

\section{Conclusions}

In this paper, we investigated the ballistic current flow in smooth graphene pn
junctions. Comparing numerically calculated current densities with
semi-classical trajectories, we demonstrated that the current flow in these
devices can be understood largely in terms of gradient-index optics. These
semi-classical trajectories are an efficient tool to estimate the current flow
paths in nano-electronic devices.

In straight pn junctions, we confirmed the validity of a generalized Snell's law
and reported additionally interference effects between the incident and
reflected electron waves, see Figures~\ref{fig:4} - \ref{fig:6}. Forbidden
regions emerge in smooth junctions, see \fig{3}. The current is reflected at the
interface of these regions, except for the normally incident electrons due to
Klein tunneling. Such smooth pn junctions can be used to generate narrow
parallel electron beams.

In circular pn junctions with an abruptly changing profile, the part of the
current density that enters the junction is focused onto a single point, which
agrees with a caustic of the semi-classical trajectories, see \fig{8}. When the
profile of the junction is smoothed, a circular junction with a gradually
changing refractive index is obtained.The semi-classical trajectories agree
qualitatively with the quantum current density but an energetically forbidden
region intensifies the interference both outside and inside the junction, see
\fig{10}. This strong interference in smooth circular pn junction leads, for
specific parameters, to interesting patterns such as whispering gallery modes,
see \fig{11}. Finally, we demonstrated in \fig{13} that smooth circular nn' and
pp' junctions act as beam splitters and converging lenses, respectively. In
\fig{15}, we proved the feasibility of realizing Luneberg and Maxwell lenses in
graphene.

We are confident that the presented broad variety of different properties of
smooth graphene pn junctions will stimulate gradient-index optics experiments in
graphene. Our findings contribute to the overall understanding of the local
current flow in graphene and may eventually lead to new nano-electronic devices.

\section{Acknowledgments}
EPR gratefully acknowledges a CONACYT graduate scholarship. EPR thanks N. Szpak
and D. Wolf for their hospitality at the University Duisburg-Essen (Germany),
where part of this research was performed. EPR, YBO and TS gratefully
acknowledge funding from CONACYT Proyecto Fronteras 952, CONACYT Proyecto
A1-S-13469, and UNAM-PAPIIT IA103020. NS gratefully acknowledges funding by the
Deutsche Forschungs\-gemeinschaft (DFG, German Research Foundation) -- Project
278162697 -- SFB 1242.

\bibliography{giopt}

\end{document}